# The carbon footprint of large astronomy meetings

Leonard Burtscher[1], Didier Barret[2], Abhijeet P. Borkar[3], Victoria Grinberg[4], Knud Jahnke, Sarah Kendrew[5], Gina Maffey[6], Mark J. McCaughrean[7]

**The annual meeting of the European Astronomical Society took place in Lyon, France, in 2019, but in 2020 it was held online only due the COVID-19 pandemic. The carbon footprint of the virtual meeting was roughly 3,000 times smaller than the face-to-face one, providing encouragement for more ecologically minded conferencing.**

The scientific evidence that we live in a climate emergency calls for urgent action[8]. As a society, we are collectively failing to live within our environmental boundaries[9] with possibly catastrophic consequences for human civilisation[1]. The time to address these issues is now[1, 10]. The United Nations Emissions Gap Report from 2019 states that each year a global reduction of emissions of 7.6% is required to limit the average global temperature rise to 1.5°C[3] - the target that was outlined in the Paris agreement in 2016[11]. At the current rate of emissions, we will exceed the 'carbon budget' to meet this goal within the next eight years[12].

While ultimately system change is required to solve the climate crisis, there is also the responsibility of individuals to reduce our emissions. This applies in particular to astronomers who rely heavily on fossil fuel energy for, e.g. computation, telescope operation and travel[13,

[1] Leiden Observatory, PO Box 9513, 2300 RA Leiden, The Netherlands
[2] Centre National de la Recherche Scientifique, Institut de Recherche en Astrophysique et Planétologie, Université de Toulouse, 9 Avenue du colonel Roche, BP 44346, F-31028, Toulouse, Cedex 4, France
[3] Astronomy Institute of the Czech Academy of Sciences, Boční II 1401, 141 00 Prague, Czech Republic.
[4] Institut für Astronomie und Astrophysik, Universität Tübingen, Sand 1, 72076 Tübingen, Germany
[5] European Space Agency, Space Telescope Science Institute, 3700 San Martin Drive, Baltimore MD 21218, USA
[6] JIVE (The Joint Institute for Very Long Baseline Interferometry European Research Infrastructure Consortium), Oude Hoogeveensedijk 4, 7991 PD Dwingeloo, The Netherlands.
[7] European Space Agency, ESTEC, Postbus 299, 2200 AG Noordwijk, The Netherlands
[8] Lenton, T.M., Rockström, J., Gaffney, O., et al. 2019. Climate tipping points - Too risky to bet against. *Nature*. DOI: 10.1038/d41586-019-03595-0.
[9] Steffen, W., Richardson, K., Rockström, J., et al. 2015. Planetary boundaries: Guiding human development on a changing planet. *Science.* DOI: 10.1126/science.1259855.
[10] United Nations Environment Programme (2019). Emissions Gap Report 2019. UNEP, Nairobi. http://www.unenvironment.org/emissionsgap
[11] *The Paris Agreement was adopted on 12 December 2015 at the twenty-first session of the Conference of the Parties to the United Nations Framework Convention on Climate Change held in Paris from 30 November to 13 December 2015.*
[12] IPCC Special Report, Global Warming of 1.5°C: https://www.ipcc.ch/sr15/
[13] Matzner C.D., Cowan N.B., Doyan R., et al. 2019. Astronomy in a low Carbon Future. https://arxiv.org/abs/1910.01272

[14,15],. To future-proof astronomy, we must recognise impending environmental change, financial uncertainties and the need for moral introspection, which threaten to hinder the continuation of the discipline. At the same time, the advancement and sharing of knowledge in general is becoming even more vital as we face a global threat.

## EWASS 2019 equivalent emissions

Conferences are a vital element of astrophysical research and collaboration, but the air travel often connected with face-to-face conferences is a major source of concern. Following last year's annual European Astronomical Society meeting in Lyon (EWASS2019), we conducted a short survey among participants who had agreed to receive such communication via email (719 out of 1240), to estimate the current, collective carbon emissions generated through travel by attendees. In establishing this initial estimate it was hoped that guidance could be developed to reduce future travel-related emissions. The anonymous questionnaire was very simple and only asked for the participants' origin and final destination and their main mode of transport. After two weeks we had collected 267 (22% of all participants) valid responses.

Just over two thirds of the respondents (66.9%) indicated that they arrived in Lyon by airplane, 27.8% arrived by train and the remaining 5.3% used other means of transport such as car, bus, metro, bike or by foot. 86.5% returned directly to their origin after the conference using the same means of transport. Of those who did not, the modal split was similar to the inbound journey.

We computed the $CO_2$-equivalent emissions (CO2eq) associated with every plane or train trip using the online Travel Footprint Calculator[16] with its default settings and we refer to the accompanying paper[17] for a discussion on the pitfalls of the methods used in these calculations (e.g. assumptions about the Radiative Forcing Index). For car trips, we used Google Maps to compute the shortest road distance and assumed emissions of 110 g/km[18]. The result of this computation is shown in Fig. 1.

[14] Williamson K., Rector T.A., and Lowenthal J. 2019. Embedding Climate Change Engagement in Astronomy Education and Research. arXiv:1907.08043
[15] Portegies Zwart, S. (in this issue of Nat. Astr.)
[16] https://travel-footprint-calculator.irap.omp.eu
[17] Barret, D.: "Estimating, monitoring and minimizing the travel footprint associated with the development of the Athena X-ray Integral Field Unit", Experimental Astronomy, 49, 3, 183-216 (2020)
[18] Average CO_2 emissions of new cars, registered in France as of 2018, https://ccfa.fr/wp-content/uploads/2018/07/fiche-pays-ccfa_co2.pdf

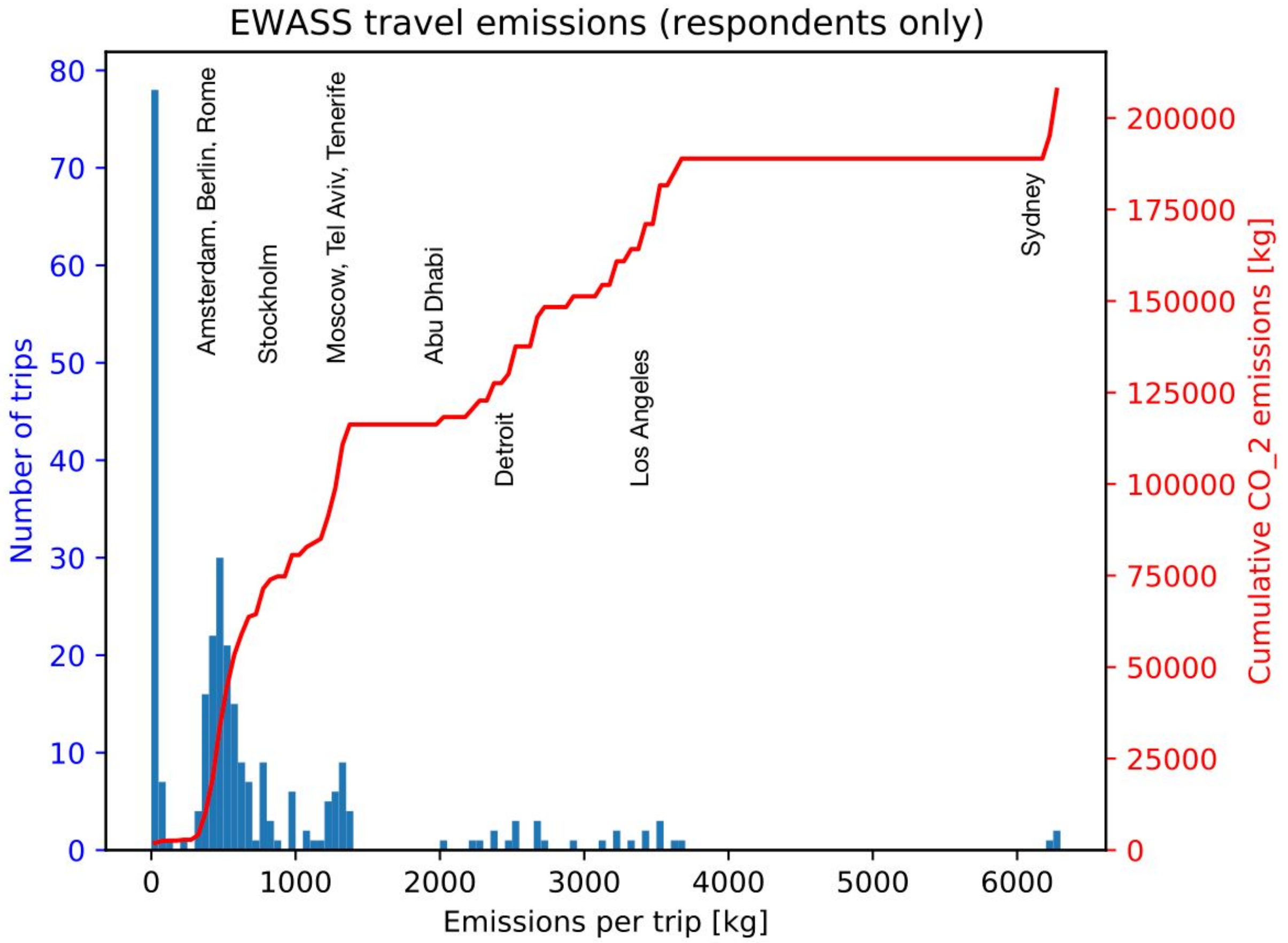

*Fig. 1: Histogram of CO2eq emissions per trip (histogram and left axis) and cumulative emissions (red line and right axis). Some example destinations are indicated for reference. Note that these numbers refer to respondents only (~ 22% of all participants).*

The majority of trips (~80%) produced CO2eq emissions of less than 1000 kg per trip. Conversely, the intercontinental flights (~ 10% of all trips) produced 50% of the total emissions of respondents.

In addition we also looked up the equivalent train travel time for all European travellers who arrived by plane, except for those where a train connection does (currently) not exist. This resulted in 159 theoretically possible train connections. 135 (85%) of these would take less than 24 hours; 114 (72%) less than 15 hours. We show a histogram of the latter together with the (looked up) train travel times for those 74 respondents who actually journeyed by train in Fig. 2.

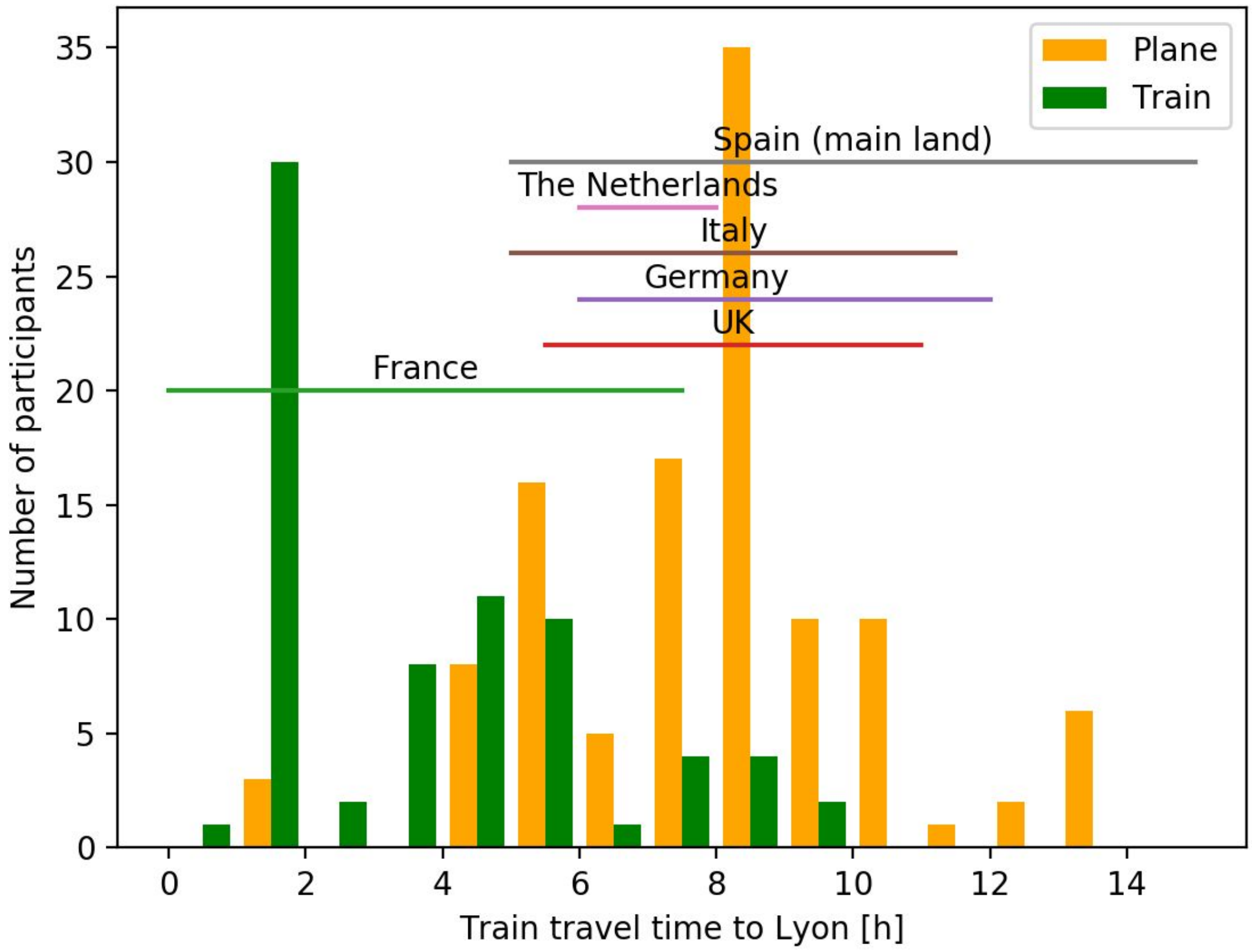

*Fig. 2: Actual or hypothetical train travel time to Lyon for all European participants who actually traveled by train (green) or could have traveled by train but took the plane (orange). Train travel times for the six countries with most participants are indicated as horizontal bars. Note that these numbers refer only to those participants who filled the survey (~ 22% of all participants).*

From additional data shared by the EAS, we know that ~ 84% of all participants (not just survey respondents) came from European destinations which are almost all accessible by train. The majority of EWASS 2019 participants could therefore have traveled to Lyon by train if they could have spent some more (working) time for the travel. It is important to note that a two hour flight usually converts into at least four hours of door-to-door travel time of which only a small fraction of time can be used for productive work. On the other hand, the median train travel time for those 85% of participants who could have traveled by train, was eight hours. Most of this time can be used productively for work, especially if traveling in modern trains that are equipped with power sockets and reliable WiFi connections. If a full week of conferencing (Monday, 9 am till Friday 5 pm) is to be combined with train travel, however, this requires flexibility for the participants to spend weekend days for travel, which may not always be possible. Another obstacle for choosing the train instead of the plane are the often higher costs of train tickets compared to equivalent airfares. This does not reflect the real costs of both modes of transport, however, since air fares do not include VAT or taxes on energy, and also do not usually include compensation for the significant environmental damage that airplanes create. On the other hand, some universities, including the Dutch universities of Leiden and Groningen and the Belgian university of Ghent, already have environmental-friendly travel policies in place that require researchers to take the train for connections that take less than six hours and recommend to take the train for connections that take up to eight hours.

Lastly, we estimated the total travel emissions associated with EWASS 2019 by correcting, country-per-country, for incompleteness in our survey responses. For example, to compute the total travel emissions for all Dutch participants, we computed the average emissions of Dutch respondents for both their outbound and return trip (673 kg) and multiplied this by the total number of participants from this country (78). This assumes a similar modal split (plane vs. train) for those participants who did not fill in the travel survey. The total travel emissions for all 1240 participants is then an estimated 1855 t CO2eq. This is of the same order of magnitude as the **yearly** emissions of a medium-sized astronomical research institute (e.g. the emissions of the MPIA in Heidelberg[19] were 2726 t CO2eq in 2018). The average per-capita travel emissions for EWASS 2019 (1.5 t) are also comparable to the **annual** emissions of developing countries (e.g. India: 1.8 t[20]). It gives a sense of the scale of the problem and the responsibility that we have that the **additional** emissions burden involved in one average astronomer participating in a single conference in Europe is roughly equal to the average total annual per capita emissions in India.

We note that our brief analysis does not include other emissions / waste associated with single-use items (such as plastic bottles, conference booklets, other give-aways), extra food production, energy for the building etc. We estimate that these factors are only a small fraction of the travel-related emissions, however (e.g. based on K. Jahnke's analysis of the carbon footprint of the Max-Planck-Institute for Astronomy[21]).

## EAS 2020 equivalent emissions

We can also compare the EWASS 2019 travel emissions to EAS 2020 -- the first online-only version of the European Astronomical Society's annual meeting: With 1777 participants, EAS 2020 was the largest online astronomy meeting to date – a change that impacted not only carbon emissions, but also diversity and inclusivity, as a forthcoming article will illustrate.

681/1777 participants (38%) answered the EAS 2020 exit survey and of those, 57% said they would have taken the plane if they had travelled to Leiden. This number is lower than last year's (67%) which may be connected to the fact that Leiden (via Rotterdam or Utrecht) is well connected to the high-speed rail network (e.g. direct connections to Paris, Brussels, London, Cologne, Frankfurt). Nevertheless, this result may also give reason to hope that astronomers are becoming more aware of the environmental damage that flying causes.

---

[19] see K. Jahnke et al. in the same issue.
[20] Muntean, M. et al.: "Fossil CO2 emissions of all world countries - 2018 Report", Publications Office of the European Union; DOI: 10.2760/30158, https://ec.europa.eu/jrc/en/publication/fossil-co2-emissions-all-world-countries-2018-report
[21] see K. Jahnke et al. in the same issue.

Of course EAS 2020 was not entirely carbon-neutral either, since virtual conferencing consumes a considerable amount of electrical energy both at the end user's site and for running the network infrastructure (and servers). A simple estimation can be made based on the fraction of people joining each of the five days of the conference (~80%, based on the exit survey), the number of hours people were online per day (5.5 hours, based on the exit survey), the data rate of a Zoom webinar (1.2 Mbps downstream[22]) and the electricity use per Gigabyte transferred over the internet (0.06 kWh/GB for 2015[23]). In addition, electrical energy is required for running the participants' laptops (~ 30 W) and the Zoom server itself. For the latter we estimate that a single 24-core Xeon machine would suffice which consumes approximately 300 W of electrical power. The total electrical energy consumption for EAS 2020 is then 1173 kWh (laptops), 1263 kWh (network) and 15 kWh (Zoom servers). With the CO2eq emission intensity for electricity generation (240 g CO2eq/kWh[24]) we arrive at a total carbon footprint for EAS 2020 of 588 kg – roughly the emissions of a **single** return trip by airplane from Liverpool to Lyon.[25]

We conclude that the internet-related emissions of EAS 2020 were insignificant compared to the travel-related emissions alone of EWASS 2019. This is comparable to other recent estimates for large international conference, e.g. a virtual annual meeting of the American Geophysical Union was calculated to emit less than 0.1% of the travel emissions of the real AGU 2019 meeting[26].

The emerging picture is that there is a real opportunity for future EAS meetings to adopt practices that provide a range of attendance possibilities for participants, which promote a more sustainable, accessible and diverse meeting concept for the growing international community.

Acknowledgements: LB would like to acknowledge discussions and feedback on a draft article from Maarten Baes, Andreas Burkert, Roger Davies, Lex Kaper, Sara Lucatello, Simon Portegies Zwart, Huub Röttgering, and Alexandra Schouten-Voskamp.

Statement on competing interests: LB was a member of the local organising committee of the EAS 2020 meeting. MMcC was a member of the scientific organising committee of the EWASS 2019 meeting. The authors declare no competing interests, however.

---

[22] Zoom bandwidth requirements: https://support.zoom.us/hc/en-us/articles/201362023

[23] Aslan, Mayers, Koomey, France: "Electricity Intensity of Internet Data Transmission – Untangling the Estimates", Journal of Industrial Ecology, 22, 4, 785 (2017), https://onlinelibrary.wiley.com/doi/pdf/10.1111/jiec.12630

[24] Extrapolating the historical figures (until 2016) at https://www.eea.europa.eu/data-and-maps/indicators/overview-of-the-electricity-production-2/assessment-4 to 2020.

[25] Note that this estimation does not take into account electricity consumption for routers/modem at home as well as other home electricity use such as lighting/heating/air condition. We assume that these facilities would anyhow be running. Since laptops are normally also in use during working hours, the number above can therefore be seen as a conservative estimate of the added emissions of joining EAS 2020 compared to working in the home office.

[26] Klöwer, Hopkins, Allen, Higham: "An analysis of ways to decarbonize conference travel after COVID-19", Nature 583, 356-359 (2020)

# Computations

## Estimation of carbon emissions of EAS 2020

### Network-related emissions

5 days * 80% participation per day * 1777 participants * 5.5 hours online per day * 1.2 Mbps * 3600s/h * ⅛ byte/bit * 1/1000 MB/GB * 0.06 kWh/GB * 0.24 kg / kWh = 303 kg CO2eq

### Laptop-related emissions

5 days * 80% participation per day * 1777 participants * 5.5 hours online per day * 30 W * 0.24 kg / kWh = 281 kg CO2eq

### Zoom-server related emissions

5 days * 10 hours per day * 300 W * 0.24 kg / kWh = 3.6 kg CO2eq